\documentclass[prc,aps,showpacs,showkeys,groupedaddress,floatfix,superscriptaddress]{revtex4-1}
\usepackage{epsfig}
\usepackage{bm}% bold math
\usepackage{graphics}
\usepackage{graphicx}
\usepackage{epsfig}
\usepackage{amssymb}
\usepackage{amsmath}
\usepackage{color}
\usepackage{natbib}
\usepackage{hyperref}

\begin{document}

\title{Interaction of the magnetic quadrupole moment of a non-relativistic particle with an electric field  in the background of screw dislocations with a rotating frame}

\author{Bekir Can L\"{u}tf\"{u}o\u{g}lu}
\email{Corresponding author : bclutfuoglu@akdeniz.edu.tr}

\affiliation{Department of Physics, Akdeniz University, Campus 07058 Antalya, Turkey}
\affiliation{Department of Physics,  University of Hradec Kr\'{a}lov\'{e}, Rokitansk\'{e}ho 62, 500\,03 Hradec Kr\'{a}lov\'{e}, Czechia}

\author{Jan K\v{r}\'{i}\v{z}}
\email{jan.kriz@uhk.cz}
\affiliation{Department of Physics,  University of Hradec Kr\'{a}lov\'{e}, Rokitansk\'{e}ho 62, 500\,03 Hradec Kr\'{a}lov\'{e}, Czechia}

\author{Soroush Zare}
\email{soroushzrg@gmail.com}
\affiliation{Department of Physics,  Shahrood University of Technology, P.O. Box 3619995161-316, Shahrood, Iran}

\author{Hassan Hassanabadi}
\email{1349hha@gmail.com}
\affiliation{Department of Physics,  University of Hradec Kr\'{a}lov\'{e}, Rokitansk\'{e}ho 62, 500\,03 Hradec Kr\'{a}lov\'{e}, Czechia}
\affiliation{Department of Physics,  Shahrood University of Technology, P.O. Box 3619995161-316, Shahrood, Iran}

\date{\today}

\begin{abstract}		
In this study, we considered a moving particle with a magnetic quadrupole moment in an elastic medium in the presence of a screw dislocation. We assumed a radial electric field in a rotating frame that leads a uniform effective magnetic field perpendicular to the plane of motion. We solved the Schr\"odinger equation to derive wave and energy eigenvalue functions by employing analytical methods for two interaction configurations:  in the absence of potential and in the presence of a static scalar potential. Due to the topological defect in medium, we observed a shift in the angular momentum quantum number which affects the energy eigenvalues and the wave function of the system.
\end{abstract}

\maketitle

%\newpage

\section{Introduction}

In recent decades, the studies on the defects in solid continua considering the theory of general relativity have gained great attention. According to these studies, the deformations of the elastic medium due to a topological defect in continuous media can be related to the differential geometry \cite{Vilenkin}. For example,  Katanaev \emph{et al.} gave a theoretical description of a continuous distribution of deformations such as disclinations and dislocations according to the Riemann-Cartan geometry \cite{Katanaev}. Puntigam \emph{et al.} explored the Voltera process and presented a geometrical description of the topological defect via a cosmic string \cite{Puntigam}.  De Assis \emph{et al.} investigated the Berry's phase in a chiral cosmic string background \cite{ek1}. Furtada \emph{et al.} examined the curvature effects on the dynamics of the particles in a two-dimensional quantum dot \cite{ek2}. Bakke \emph{et al.} discussed the neutral particle's Landau quantization in the presence of the topological defects \cite{ek3}.

The elastic deformation in a solid can occur as a linear topological defect caused by mechanical working \cite{Solyom}. The screw dislocation and the spiral dislocation are called as the models of the linear topological defects. These kinds of topological defects have widely been used in the literature with the intention of studying quantum dynamics in a medium with distortion. For instance, Dantas \emph{et al.}  examined the quantum dynamics of a rotating electron/hole in an existing screw dislocation within a confined ring potential energy \cite{DantasPLA2015}.  Furtado \emph{et al.} considered an external magnetic field and explored the dynamics of the electrons in a medium with the presence of screw dislocation  \cite{FurtadoEPL1999}. Da Silva   \emph{et al.}  considered a  non-relativistic particle in an elastic medium and discussed the  Aharonov-Bohm-type effects in the background of a distortion of a vertical line into a vertical spiral \cite{daSilvaEPJC2019}. Some of the other recent researches that have been carried on can be found in Refs. \cite{Bausch,Netto,Furtado1994,zareEPJP, screw1, screw2, screw3, screw4, screw5, screw6, screw7}.

Over the past few years, many researchers examined the interaction between curved spacetime and quantum fields. For example, Bezerra explored the wave functions and the energy spectra of scalar and spinor quantum particles in the spacetime background which is assumed by generated by a chiral cosmic string \cite{BezerraJMP1997}. Marques \emph{et al.} examined the hydrogen atom's energy spectra by solving the Dirac equation in the background spacetimes which are assumed to be induced from a cosmic string and global monopole \cite{MarquesPRD2002}. Fonseca \emph{et al.} considered an atom with a magnetic quadrupole moment and investigated the rotating effect while the atom is interacting with an electric field radially in a two-dimensional quantum ring potential \cite{FonsecaEPJP2016}. In one of two very recent papers, Hassanabadi \emph{et al.} solved the Schr\"odinger equation analytically for a moving particle in a rotating frame which interacts with a radial external electric field by its own magnetic quadrupole moment  \cite{HassanabadiAP2020}.  In the other one, Vieira \emph{et al.} considered a moving particle with a magnetic quadrupole moment and investigated the bound state solutions for the electromagnetic field interactions within a uniformly rotating frame by using a semiclassical approach \cite{Vieira2020}. Furthermore, the rotating effects have been surveyed by researching in many subjects of physics, such as the Sagnac effect \cite{Sagnac1, Sagnac2, Post} , the relativistic quantum mechanics \cite{Hbakke3, rot1}, the quantum rings \cite{rot4, rot2, Dantas2015}, the quantum spintronics \cite{Matsuo2011, rot6}, and the quantum Hall effect \cite{Fischer}. It should be noted that rotating effects have been studied in the Dirac \cite{Iyer} and scalar field theories \cite{Letaw, Konno}. Also rotating effects in the context of the condensed matter physics, in particular in Bose-Einstein condensation in ultracold diluted atomic gases, have been investigated \cite{Hu}. For further reading, we remark some of the other studies on the subject in \cite{FurtadoPLA1994,Aharonov,He1993,Wilkens1994,FonsecaJCP2016,FonsecaAP2015,deMontigny2018, rot3, rot5, rot7, FonsecaJMP2017}.

In this study, we examine the interaction of a radial electric field with the magnetic quadrupole moment of a moving particle system by solving the corresponding Schr\"odinger equation in the presence of a screw dislocation in an elastic medium. We investigate the effect of rotation on this non-relativistic system by considering a rotating frame with a constant angular velocity since rotating frame effects are giving rise the geometric quantum phases such as the Aharonov-Carmi phase \cite{AharonovCarmi1973, AharonovCarmi1974} and the Mashhoon effect \cite{Mashhoon}.

This paper is organized as follows. In Sect. \ref{sec2}, we consider a non-relativistic moving particle with a magnetic quadrupole moment in an elastic medium in the presence of screw dislocation and derive the Schr\"odinger equation to examine quantum dynamics while the moving particle is in a rotating frame and interacts with the external electric field.  In Sect. \ref{sec3}, we solve the Schr\"odinger equation and obtain the energy eigenvalue and wave functions for two configurations: in the absence of potential energy, and in the presence of a static scalar potential energy by using the Nikiforov-Uvarov (NU) method. Finally, in Sect. \ref{Conc} we conclude the manuscript.

\section{Magnetic quadrupole moment of a moving particle in the presence of a screw dislocation\label{sec2}}

Before describing the dynamic of a non-relativistic scalar particle with a magnetic quadrupole moment that is located in a region which possesses a uniform effective magnetic field and a external electric field, we would like to touch on a screw dislocation in an elastic medium. It is worth noting that dislocations as the topological defects in a lattice can be classified as spiral dislocations and screw dislocations. When a considerable number of bonds are broken between atoms along the line a crystalline material, so part of that lattice is offset by a bit relative to the other part of the lattice.  In fact, it causes distortion in a crystalline, so-called the screw dislocation. We employ natural units, $\hbar=\mathrm{c}=1$, and express the line element which is associated with a screw dislocation oriented along the $z$-axis in the spacetime with the kind of topological defects.Indeed, this topological defect that carries torsion (but no curvature) can be described by the following line element \cite{DantasPLA2015,FurtadoEPL1999,Marques2001,Netto}:
\begin{equation}\label{le1}
\mathrm{d}s^2=-\mathrm{d}t^2+\mathrm{d}\rho^2+\left(\mathrm{d}z+\beta\mathrm{d}\varphi\right)^2+\rho^2\mathrm{d}\varphi^2,
\end{equation}
where $\phi$, and $\rho$ are the azimuthal and radial coordinates, respectively. The ranges of the line coordinate elements are $0<t<\infty$, $0<\rho=\sqrt{x^2+y^2}$, $0<\varphi<2\pi$, and $-\infty<z<\infty$. The covariant metric tensor $\mathrm{g}_{ij}$ associated with the spatial part of the line element Eq. \eqref{le1} is given by
\begin{equation}
\mathrm{g}_{ij}=\begin{pmatrix}
1&0&0\\
0&\rho^2+\beta^2&\beta\\
0&\beta&1
\end{pmatrix}, \qquad i,j= \rho, \varphi,z.
\end{equation}
Here, $\beta$ is a parameter associated with the Burgers vector $\mathrm{b}_{z}$ by $\beta=\frac{\mathrm{b}_{z}}{2\pi}$. It is worth noting that the Burgers vector is an important concept in linear defects in order to describe the character of a dislocation in a crystalline material. Such as, the Burgers vector related to a screw dislocation is contained in the defect region while the Burgers vector related to a spiral dislocation is found to be perpendicular to the defect region \cite{Puntigam}.

Next, let us present the magnetic multipole expansion since our focus lies on the magnetic quadrupole polarizability tensor. Under an external magnetic field, $\vec{\mathrm{B}}$ (whose components 	are determined by $\mathrm{B}_i$), the magnetic potential energy of a localized distribution of moving charged particles or currents in a medium in the rest frame of the particle can be written as
\begin{equation}\label{expEnrLocDistrib}
U_B=-\sum_{i}\mathrm{B}_{i}(0) \mu_i -\frac16\sum_{i,j}\frac{\partial \mathrm{B}_{i}(0)}{\partial x_{j}}\mathrm{M}_{ij}-\frac{1}{24}\sum_{i,j,k}\frac{\partial^2 \mathrm{B}_{i}(0)}{\partial x_{j}\partial x_{k}}\mathrm{R}_{ijk}+\dots,
\end{equation}
where $\vec{\mu}$ is the total magnetic dipole moment defined  by
\begin{equation}\label{MagDipol}
\mu_i\equiv\frac{1}{2} \sum_{j,k}\mathcal{\epsilon}_{ijk} \int x_{j}\mathrm{j}_k \,\,\mathrm{d}^{3}x.
\end{equation}
Here, $\mathcal{\epsilon}_{ijk}$ is the Levi-Civita tensor, $x_{j}$ and $j_{\ell}$ are components of the position, $\vec{\mathrm{r}}$, and current-density distribution vectors, $\vec{\mathrm{J}}$, respectively. In the second term of the right-hand side of Eq. \eqref{expEnrLocDistrib}, $\mathrm{M}_{ij}$, is the magnetic quadrupole moment tensor, and it is defined in the form of
\begin{equation}\label{3}
\mathrm{M}_{ij}\equiv \sum_{k,\ell} \mathcal{\epsilon}_{ik\ell} \int \left(3x_{j}x_{k}-\mathrm{r}^2\delta_{jk}\right)j_{\ell}\mathrm{d}^3 x,
\end{equation}
in which, $x_{j}$ is the $j^{th}$ component of the position vector $\vec{\mathrm{r}}$, such that the magnitude of the $\vec{\mathrm{r}}$ is denoted by $\mathrm{r}$, the $j_{\ell}$ is denoted as the components of the $\vec{\mathrm{J}}$, and  the Kronecker delta function is indicated by $\delta_{jk}$. Meanwhile, the tensor $\mathrm{M}_{ij}$ is traceless, that is, $\sum_{i} \mathrm{M}_{ii}=0$, it is a symmetric tensor as well. As regards the third term of the right-hand side of Eq. \eqref{expEnrLocDistrib}, the magnetic octupole moment tensor $\mathrm{R}_{ijk}$ is defined as	
\begin{equation}\label{5}
\mathrm{R}_{ijk}\equiv 6 \sum_{\ell,m} \epsilon_{i\ell m}\int x_{\ell} x_{j} x_{k} j_{m} \mathrm{d}^3 x.
\end{equation}
Hereafter, we use the natural units, $\mathrm{q}=\mathrm{c}=\hbar=1$. We follow the references \cite{FonsecaJCP2016,FonsecaAP2015,Radt1970,Dmitriev1994,FonsecaJMP2015}, and consider a non-relativistic moving particle that possesses only a magnetic quadrupole moment. Note that, in the frame of the moving particle,  the particle experiences a different magnetic field, let us call it by $\vec{\acute{\mathrm{B}}}$, which can be obtained via the Lorentz transformation as $\vec{\acute{\mathrm{B}}}=\vec{\mathrm{B}}- \vec{v} \times \vec{E}$ (up to $O(v^2)$ meanwhile, the velocity vector of a charged particle is given by $\vec{v}$) terms. Here,  $\vec{\mathrm{E}}$ and $\vec{\mathrm{B}}$ are  the electric and magnetic fields in the laboratory frame, respectively. Thus, the lagrangian of this system can be written as
\begin{equation}\label{Lagr1}
\mathrm{L}= \frac12 \mathrm{m} v^2+\sum_{ij}\mathrm{M}_{ij}\frac{\partial \acute{\mathrm{B}}_{i}}{\partial x_{j}},
\end{equation}
in which $\mathrm{m}$ is mass of a neutral particle. Note that the second term of the right-hand side of Eq. \eqref{Lagr1} has been supposed as the potential energy of a neutral particle with a magnetic quadrupole moment (as defined in Refs. \cite{FonsecaJCP2016,FonsecaAP2015,Radt1970,Dmitriev1994,FonsecaJMP2015}). Furthermore, $\mathrm{M}_{ij}$ is a symmetric and traceless tensor which presents the magnetic quadrupole moment tensor \cite{Radt1970,Dmitriev1994} and $\vec{\acute{\mathrm{B}}}$ is the magnetic field whose components are determined by $\acute{\mathrm{B}}_{j}$. Now, let us rewrite the lagrangian Eq. \eqref{Lagr1}, according to $\vec{\acute{\mathrm{B}}}$, as
\begin{equation}\label{Lagr2}
\mathrm{L}=\frac12 \mathrm{m} v^2+\vec{\mathrm{M}}.\vec{\mathrm{B}}+\vec{v}.\left(\vec{\mathrm{M}}\times\vec{\mathrm{E}}\right),
\end{equation}
in which
$\vec{\mathrm{M}}$, according to Eqs. \eqref{Lagr1} and \eqref{Lagr2}, is a vector whose components are defined by
\begin{equation}\label{ComM}
\mathrm{M}_{i}=\sum_{j} \mathrm{M}_{ij}\partial_{j},
\end{equation}
now, let us present the non-null components of the magnetic quadrupole moment tensor $\mathrm{M}_{ij}$ as
\begin{equation}\label{NNCMQMT}
\mathrm{M}_{\rho z}=\mathrm{M}_{z \rho}=\mathrm{M},
\end{equation}
where $\mathrm{M}$ is a positive constant. Then, we obtain the  canonical momentum \cite{Landau1980}
\begin{equation}
\vec{\mathrm{p}}=\mathrm{m} \vec{v}+\vec{\mathrm{M}}\times\vec{\mathrm{E}},
\end{equation}
to introduce the Hamiltonian of this system with
\begin{equation}
H=\frac{1}{2\mathrm{m}}\left|\vec{\mathrm{p}}-\vec{\mathrm{M}}\times\vec{\mathrm{E}}\right|^2-\vec{\mathrm{M}}.\vec{\mathrm{B}}.
\end{equation}
where $\vec{\mathrm{p}}=-i\vec{\nabla}$ is the momentum operator.  If we compare the above momentum with $\vec{\mathrm{p}}-\mathrm{q}\vec{\tilde{\mathrm{A}}}^{\mathrm{eff}}$, we conclude that $\vec{\tilde{\mathrm{A}}}^{\mathrm{eff}}=\vec{\mathrm{M}}\times\vec{\mathrm{E}}$
 and $V^{\mathrm{eff}}=-\vec{\mathrm{M}}.\vec{\mathrm{B}}$, as in \cite{FonsecaJMP2015,Landau1977}. Note that we do not have any external magnetic field in this configuration; therefore, we get $V^{\mathrm{eff}}=0$.

Thus, we express the Schr\"odinger equation for a quantum particle with a magnetic quadrupole moment in a rotating frame in the presence of a static scalar potential energy $\mathrm{V}\left(\rho\right)$ \cite{DantasPLA2015,HassanabadiAP2020, Vieira2020}.
\begin{equation}\label{Sch1}
\left(\frac{\vec{\Pi}^2}{2\mathrm{m}}-\vec{\Omega}.\vec{\mathrm{L}}+\mathrm{V}\left(\rho\right)\right)\Psi\left(t,\vec{r}\right)=i\partial_{t}\Psi\left(t,\vec{r}\right).
\end{equation}
Here, $\vec{\mathrm{L}}$ is the angular momentum operator and $\vec{\Omega}$ is the constant angular velocity which is taken as $\vec{\Omega}=\omega\hat{z}$, since we assume the rotating frame along the $z$-axis.  In references \cite{DantasPLA2015,FurtadoEPL1999,daSilvaEPJC2019,BezerraJMP1997}, it is shown that the presence of the torsion influenced the angular momentum. Particularly, it is noted that the $z$-component of the angular momentum, $\hat{\mathrm{L}}_{z}=-i\partial_{\varphi}$, is being modified by the torsion effect with an additional term.

Before we derive the term associated with the rotating frame, we recall the definition of the angular momentum, $\vec{\mathrm{L}}=\vec{\mathrm{r}}\times\vec{\Pi}$, where $\vec{\mathrm{r}}=\rho\hat{\rho}$ and $\rho$ is the radial coordinate of the cylindrical coordinate system. Note that $\hat{\rho}$ is the unit vector that points the radial direction. We use the definition of the generalized momentum operator
\begin{equation}\label{geneM}
\vec{\Pi}=\vec{\mathrm{p}}-\vec{\mathrm{M}}\times\vec{\mathrm{E}}.
\end{equation}
Solving Eq. \eqref{Sch1} begins by finding $\vec{\Pi}^2\Psi\left(t,\vec{r}\right)$, therefore, it can be written as follows
\begin{equation}\label{term1Sch1}
	\hat{\Pi}^2\Psi\left(t,\vec{r}\right)=\left(-\nabla^2+2i\left(\vec{\mathrm{M}}\times\vec{\mathrm{E}}\right).\vec{\nabla}
+\left(\vec{\mathrm{M}}\times\vec{\mathrm{E}}\right)^2\right)\Psi\left(t,\vec{r}\right).
\end{equation}
We proceed further with the introduction of the radial electric field vector $\vec{\mathrm{E}}$:
\begin{equation}\label{elecf}
\vec{\mathrm{E}}=\frac{\lambda}{2}\rho^2\hat{\rho},
\end{equation}
where $\lambda$ is a constant that denotes a non-uniform distribution of electric charges inside the non-conducting cylinder.

The selection of the electric field for the laboratory frame as given in Eq. \eqref{elecf},  let us to define a uniform effective magnetic field out of the interaction between the radial electric field and magnetic quadrupole moment in the form of
\begin{equation}\label{effmagf}
\vec{\mathrm{B}}^{\mathrm{eff}}=\vec{\nabla}\times\left(\vec{\mathrm{M}}\times\vec{\mathrm{E}}\right)=2\mathrm{M}\lambda \hat{z}.
\end{equation}
According to Eqs. \eqref{elecf} and \eqref{effmagf}, let us now rewrite Eq. \eqref{term1Sch1}  as follows
\begin{equation}
\hat{\Pi}^2\Psi\left(t,\vec{r}\right)=\left(-\nabla^2+2i\mathrm{M}\lambda\rho\vec{\nabla}_{\varphi}+\mathrm{M}^2\lambda^2\rho^2\right)\Psi\left(t,\vec{r}\right).
\end{equation}
Then, we use the  line element that is given in Eq. \eqref{le1}, to derive the Laplacian operator \cite{Arfken}. We find
\begin{equation}
\nabla^2=\frac{1}{\rho}\partial_{\rho}\left(\rho\partial_{\rho}\right)+\frac{1}{\rho^2}\left(\partial_{\varphi}-\beta\partial_{z}\right)^2+\partial_{z}^2.
\end{equation}
Therefore we conclude that the angular momentum operator is being modified via $\partial_{\varphi}\rightarrow\partial_{\varphi}-\beta\partial_{z}$ in the elastic medium with the screw dislocation.
\begin{equation}\label{modL}
\hat{\mathrm{L}}_{z}^{\mathrm{eff}}=-i\left(\partial_{\varphi}-\beta\partial_{z}\right).
\end{equation}
 Note that this operator shows the effect of the torsion on the $z$-component of the angular momentum. In Ref. \cite{Maia} it is shown that the topology of a screw dislocation that corresponds to the distortion of a circular curve into a vertical spiral makes such transformation and also showed that the effective operator of the $z$-component of the angular momentum does not commute with the Hamiltonian. Next, the effects of rotation on this nonrelativistic system are investigated by $\vec{\Omega}.\vec{\mathrm{L}}\Psi\left(t,\vec{r}\right)$ (the second term of the left-hand side of Eq. \eqref{Sch1}); thereby, it can be found as
\begin{equation}
\vec{\Omega}.\vec{\mathrm{L}}\Psi\left(t,\vec{r}\right)\equiv\omega\hat{\mathrm{L}}^{\mathrm{eff}}_{z}
\Psi\left(t,\vec{r}\right)=\left[-i\omega\left(\partial_{\varphi}-\beta\partial_{z}\right)-\omega\mathrm{M}\lambda\rho^2\right]\Psi\left(t,\vec{r}\right).
\end{equation}
To solve Eq. \eqref{Sch1}, we consider a particular solution. Since the Hamiltonian commutes with the operators $\hat{\mathrm{p}}_{z}$ and $\hat{\mathrm{L}}_{z}^{\mathrm{eff}}$, we make an ansatz and assume the wave function in the following form.
\begin{equation}\label{WF1}
\Psi\left(t,\vec{r}\right)=e^{-i\mathcal{E}t+i\ell\varphi+ikz}\psi\left(\rho\right).
\end{equation}
Here,  $\mathcal{E}$ is the energy of the system, $k$ denotes a constant that is so-called the wave number along the $z$-axis, and $\ell$ is a quantum number where $\ell=0,\pm1,\pm2,\dots$. By employing the wave-function solution in Eq. \eqref{Sch1}, we find the second-order time-independent radial Schr\"odinger equation as
\begin{eqnarray}
\Bigg[\frac{\mathrm{d}^2}{\mathrm{d}\rho^2}+\frac{1}{\rho}\frac{\mathrm{d}}{\mathrm{d}\rho}-\frac{1}{\rho^2}\left(\ell-\beta k\right)^2-k^2+2(\mathrm{M}\lambda+\mathrm{m}\omega)\left(\ell-\beta k\right)+2m\mathcal{E}-2m\mathrm{V}\left(\rho\right)-{(\mathrm{M}^2\lambda^2+2\mathrm{m}\mathrm{M}\lambda\omega)}\rho^2\Bigg]\psi\left(\rho\right)=0.\,\, \label{Sch2}
\end{eqnarray}

\section{Exact solution of the Schr\"odinger equation\label{sec3}}
In this section, we solve the derived time-independent Schr\"odinger equation for two different choices of the scalar potential energy: with the absence of the static potential energy and with the pseudo-harmonic type potential energy. In the both  solutions, we employ the Nikiforov–Uvarov (NU) method to obtain the energy eigenvalue and the wave functions. We avoid writing the NU method here to prevent unnecessary repetition. For details, one can see e.g. Appendix in Ref. \cite{deMontigny2018}.

\subsection{The first case}
\begin{figure}[!htb]
	\begin{center}
\includegraphics[scale=1]{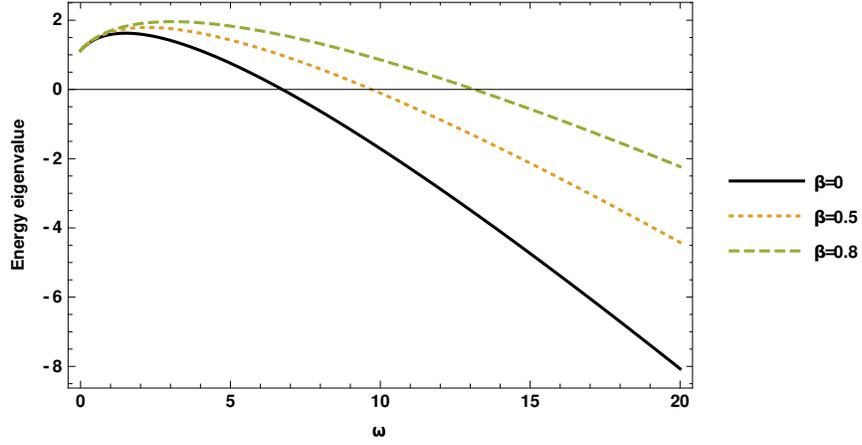}
		\caption{The behavior of energy eigenvalue functions versus the angular velocity with various Burger parameters.}\label{Figy1}
	\end{center}
\end{figure}

In this case, we solve Eq. \eqref{Sch2} in the absence of static potential energy, that is, $\mathrm{V}\left(\rho\right)=0$. We find this choice interesting since the magnetic potential energy is also absent. Thereby, this case corresponds to the moving particle's quantum dynamics in a rotating frame by considering the interaction of its magnetic quadrupole moment with the external radial electric field. We start the solution by dropping the static potential energy terms out of Eq. \eqref{Sch2}. We find
\begin{eqnarray}
&&\frac{\mathrm{d}^2\psi_{n\ell}\left(\rho\right)}{\mathrm{d}\rho^2}+\frac{1}{\rho}\frac{\mathrm{d}\psi_{n\ell}\left(\rho\right)}{\mathrm{d}\rho}
+\frac{1}{\rho^2}\bigg[-\left(\mathrm{M}^2\lambda^2+2\mathrm{m}\mathrm{M}\lambda\omega\right)\rho^4+\Big(2(\mathrm{M}\lambda+\mathrm{m}\omega)\left(\ell-\beta k\right)-k^2+2\mathrm{m}\mathcal{E}\Big)\rho^2 \nonumber \\
&-&\left(\ell-\beta k\right)^2\bigg]\psi_{n\ell}\left(\rho\right)=0.\label{Sch3}
\end{eqnarray}
We proceed by introducing a new changing variable,  $s=\rho^2$. Thus, we  rewrite Eq. \eqref{Sch3}  as
\begin{eqnarray}\label{Sch4}
&&\frac{\mathrm{d}^2\psi_{n\ell}\left(s\right)}{\mathrm{d}s^2}+\frac{1}{s}\frac{\mathrm{d}\psi_{n\ell}\left(s\right)}{\mathrm{d}s}
+\frac{1}{4s^2}\bigg[-\left(\mathrm{M}^2\lambda^2+2\mathrm{m}\mathrm{M}\lambda\omega\right)s^2+\Big(2(\mathrm{M}\lambda+\mathrm{m}\omega)\left(\ell-\beta k\right)-k^2+2\mathrm{m}\mathcal{E}\Big)s \nonumber \\
&&-\left(\ell-\beta k\right)^2\bigg]\psi_{n\ell}\left(s\right)=0.
\end{eqnarray}
Then, we compare Eq. \eqref{Sch4} with the NU equation and obtain the wave function with regard to the generalized Laguerre polynomials $\mathrm{L}^{\alpha}_{n}\left(x\right)$:
\begin{equation}\label{WF2}
\psi_{n\ell}\left(\rho\right)=\mathrm{N}_{nl}\rho^{2\alpha_{12}}e^{\alpha_{13}\rho^2}\mathrm{L}^{\alpha_{10}-1}_{n}\left(\alpha_{11}\rho^2\right),
\end{equation}
where $\mathrm{N}_{nl}$ is the normalization constant. Moreover, we find the coefficients $\alpha_{10}$, $\alpha_{11}$, $\alpha_{12}$ and $\alpha_{13}$ as follows:
\begin{equation}\label{alphas1}
\begin{split}
&\alpha_{10}=1+\left|\ell-\beta k\right|, \qquad \alpha_{11}=\sqrt{\mathrm{M}^2\lambda^2+{2}\mathrm{m}\mathrm{M}\lambda\omega},\\
&\alpha_{12}=\frac{\left|\ell-\beta k\right|}{2}, \qquad \qquad \alpha_{13}={-\frac12}\sqrt{\mathrm{M}^2\lambda^2+2\mathrm{m}\mathrm{M}\lambda\omega}.
\end{split}
\end{equation}
Then, we follow the NU method and derive the energy eigenvalue function as
\begin{equation}\label{energyc1}
\mathcal{E}_{n\ell}=\frac{k^2}{2\mathrm{m}}-\left(\frac{\mathrm{M}\lambda}{\mathrm{m}}+\omega\right)\left(\ell-\beta k\right)+{\frac1m}\left(1+2n+\left|\ell-\beta k\right|\right)\sqrt{\mathrm{M}^2\lambda^2+2\mathrm{m}\mathrm{M}\lambda\omega}.
\end{equation}
Henceforth, we discuss the behavior of energy eigenvalues $\mathcal{E}_{n\ell}$ in terms of the constant angular velocity $\omega$.  At a particular value of the angular velocity,  the energy eigenvalues become zero. This special value of the $\omega$ is called a critical angular velocity and we denote it with  $\omega_{c}$.
We obtain the critical angular velocity  as
\begin{equation}\label{omegacrit1}
	\begin{split}
		\omega_{c}=&\frac{1}{8\left(\ell-\beta k\right)^2}\left\lbrace\frac{1}{\mathrm{m}}\Big( 4k^2\left(\ell-\beta k\right) +8\mathrm{M}\lambda\left(1+2n\right)^2+16\mathrm{M}\lambda\left(1+2n\right)\left|\ell-\beta k\right|\Big)\right.\\
		& \pm\left[\frac{1}{\mathrm{m}^2}\Big( 4k^2\left(\ell-\beta k\right) +8\mathrm{M}\lambda\left(1+2n\right)^2+16\mathrm{M}\lambda\left(1+2n\right)\left|\ell-\beta k\right|\Big)^2\right.\\
		&\left.\left.+\frac{16}{\mathrm{m}^2}\left(\ell-\beta k\right)^2\Big(4\mathrm{M}^2\lambda^2\left(1+2n\right)^2-\left(k^2-2\ell\mathrm{M}\lambda\right)^2-4k\mathrm{M}\lambda\beta\left(k^2-2\ell\mathrm{M}\lambda+k\mathrm{M}\beta\lambda\right)\right.\right.\\
		&\left.\left.+8\mathrm{M}^2\lambda^2\left(1+2n\right)\left|\ell-\beta k\right|+4\mathrm{M}^2\lambda^2\left|\ell-\beta k\right|^2\Big) \right]^{1/2}\right\rbrace.
	\end{split}
\end{equation}
We choose the positive sign in the brackets  and take  $\ell=\lambda=\mathrm{m}=\mathrm{M}=1$, $k=0.5$ to perform  numerical calculations. For $\beta=0.5$ from Eq.\eqref{omegacrit1} we calculate the first four, $n=0,1,2,3$, critical angular velocities as $9.71$, $48.83$, $116.39$, and $212.39$, respectively. Then, we extend our analysis by   considering two different values of the dislocation parameter, $\beta=0$ and $\beta=0.8$. We use Eq.\eqref{energyc1} and demonstrate the behavior of the energy eigenvalue functions versus the angular velocity in Fig.~\ref{Figy1}.
It is worth noting that for $\omega>\omega_{c}$ values we find the negative energy eigenvalues, while  $\omega<\omega_{c}$  values we obtain  positive energy eigenvalues. We observe that in the absence of dislocation, for a lower value of the angular velocity,  more precisely for $\omega_c= 6.73$, the particle becomes confined. When the dislocation parameter values increases, higher angular velocity is required for the confinement.

\begin{figure}[!htb]
	\begin{center}
\includegraphics[scale=1]{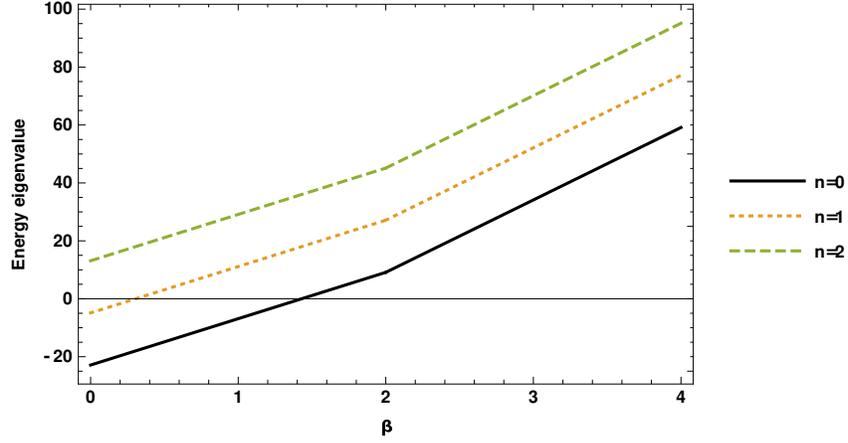}
		\caption{The behavior of energy eigenvalue functions versus the Burger parameter.}\label{Figy2}
	\end{center}
\end{figure}

Next, we investigate the variation of the energy eigenvalue versus the dislocation parameter at an arbitrary angular velocity value. We choose $\omega=20$ and for $n=0,1,2$ we present the plot of the energy eigenvalue function versus $\beta$ in Fig.\ref{Figy2}.
We observe that in the ground state, with the assigned angular velocity the particle stays in confinement up to $1.43$ disorder value. Regarding the first excited states, the confinement is valid up to $0.30$ disorder value. when we consider the second excited states, we do not observe  confinement. Besides, we observe a critical value of the Burger parameter, $\beta_c=k/l$. After this critical value, the change in the energy eigenvalue has a different rate.

 Before we examine a non-zero scalar potential energy, we would like to discuss the role of the parameters on the degeneracy of the eigenvalue function. If we consider the system does not have a dislocation and rotation,  the energy eigenvalue function reduces to
\begin{eqnarray}
\mathcal{E}_{n\ell}&=&\frac{k^2}{2\mathrm{m}}+\frac{M \lambda}{m}\big(1+2n+\left|\ell\right|-\ell\big). \label{eq30}
\end{eqnarray}
We find that  for the $\ell \geq 0$ case the energy eigenvalue becomes  independent of $\ell$ parameter, $\mathcal{E}_{n\ell}=\frac{k^2}{2\mathrm{m}}+\frac{M \lambda}{m}(1+2n)$. On the other hand, when we consider the $\ell<0$ values, the $\ell$  dependency continues. Moreover, we observe a degeneracy for any $n+\ell=n'+\ell'$ values.
\\
If we consider the system with a dislocation, $\beta \neq 0$, but without a rotation, $\omega=0$, then the energy eigenvalue function turns to the following form:
\begin{eqnarray}
\mathcal{E}_{n\ell}&=&\frac{k^2}{2\mathrm{m}}+{\frac{\mathrm{M}\lambda}{m}}\big(1+2n+\left|\ell-\beta k\right|-\left(\ell-\beta k\right)\big), \label{eq31}
\end{eqnarray}
In this case, for the $\ell-\beta k \geq 0$  values, we obtain
\begin{eqnarray}
\mathcal{E}_{n\ell} &=& \frac{k^2}{2\mathrm{m}}+\frac{M \lambda}{m}(1+2n)
\end{eqnarray}
which shows that the $\ell$ parameter dependence is lost. In the other case where $\ell-\beta k < 0$, we arrive at
\begin{eqnarray}
\mathcal{E}_{n\ell}&=& \frac{k^2}{2\mathrm{m}}+\frac{M \lambda}{m}(1+2n+2\left|\ell-\beta k\right|).
\end{eqnarray}
We observe a degeneracy in the eigenvalue function for the $n+\ell=n'+\ell'$ quantum numbers. As a result of this analysis, we conclude that the angular velocity and the quantum number $\ell$ removes the degeneracy of the  system.

It is worth mentioning that, for $k=0$, Eqs. \eqref{eq30} and \eqref{eq31} are the same to the Eq. (12) of \cite{FonsecaJCP2016} if and only if $\Omega=0$ is considered there. Furthermore, they are very similar to the Eq. (27) of \cite{FonsecaAP2015}.

\newpage
\subsection{The second case}

In this case, we consider a static scalar potential in the form of pseudo-harmonic type potential energy \cite{DantasPLA2015}.
\begin{equation}\label{sspot}
\mathrm{V}\left(\rho\right)=C_1\rho^2+\frac{C_2}{\rho^2}+C_3,
\end{equation}
It is worth mentioning that many applications have been reported by this kind of static scalar potential,  for example a quantum ring of a nanosphere \cite{Tan1996, NettoFarias2019}. It is a basic example of central potential energies that is constituted with the radial harmonic oscillator, inverse square and a constant terms.  Its coefficients $C_1$, $C_2$ and $C_3$ are real valued constants. We start the examination by substituting the considered potential energy into Eq. \eqref{Sch2}. We apply a change of the variable, $\tilde{\rho}=\rho^2$. We get
\begin{eqnarray}\label{Sch5}
&&\frac{\mathrm{d}^2\psi_{n\ell}\left(\tilde{\rho}\right)}{\mathrm{d}\tilde{\rho}^2}+\frac{1}{\tilde{\rho}}\frac{\mathrm{d}\psi_{n\ell}\left(\tilde{\rho}\right)}{\mathrm{d}\tilde{\rho}}
-\frac{1}{4\tilde{\rho}^2}\bigg[\left(\mathrm{M}^2\lambda^2+2\mathrm{m}\mathrm{M}\lambda\omega+2\mathrm{m}C_1\right)\tilde{\rho}^2-\Big(2(\mathrm{M}\lambda+\mathrm{m}\omega)\left(\ell-\beta k\right)-k^2+2\mathrm{m}\mathcal{E}-2\mathrm{m}C_3\Big)\tilde{\rho} \nonumber \\
&&+\left(\ell-\beta k\right)^2+2\mathrm{m}C_2\bigg]\psi_{n\ell}\left(\tilde{\rho}\right)=0.
\end{eqnarray}
Alike {\bf{the first case}}, we solve  Eq. \eqref{Sch5} with the NU method. We obtain the wave functions as
\begin{equation}\label{WF3}
\psi_{n\ell}\left(\rho\right)=\tilde{\mathrm{N}}_{nl}\rho^{2\gamma_{12}}e^{\gamma_{13}\rho^2}\mathrm{L}^{\gamma_{10}-1}_{n}\left(\gamma_{11}\rho^2\right),
\end{equation}
where $\tilde{\mathrm{N}}_{nl}$ is the normalization constant. From the coefficients
\begin{equation}\label{alphas2}
\begin{split}
&\gamma_{10}=1+\sqrt{2\mathrm{m}C_{2}+\left(\ell-\beta k\right)^2}, \qquad \gamma_{11}=\sqrt{2\mathrm{m}C_{1}+\mathrm{M}^2\lambda^2+2\mathrm{m}\mathrm{M}\lambda\omega},\\
&\gamma_{12}=\frac12\sqrt{2\mathrm{m}C_{2}+\left(\ell-\beta k\right)^2},\,\,\,\, \qquad \gamma_{13}=-\frac12\sqrt{2\mathrm{m}C_{1}+\mathrm{M}^2\lambda^2+2\mathrm{m}\mathrm{M}\lambda\omega},
\end{split}
\end{equation}
we derive the energy eigenvalue function as
\begin{equation}\label{energyc2}
\mathcal{E}_{n\ell}=C_{3}+\frac{k^2}{2\mathrm{m}}-\left(\frac{\mathrm{M}\lambda}{\mathrm{m}}+\omega\right)\left(\ell-\beta k\right)+\frac{1}{m}\left(1+2n+\sqrt{2\mathrm{m}C_{2}+\left(\ell-\beta k\right)^2}\right)\sqrt{2\mathrm{m}C_{1}+\mathrm{M}^2\lambda^2+2\mathrm{m}\mathrm{M}\lambda\omega}.
\end{equation}
Note that for $C_1=C_2=C_3=0$, Eq.\eqref{energyc2} reduces to Eq.\eqref{energyc1}. As we have done in {\bf{the first case}}, we derive the critical angular velocity function,  $\tilde{\omega}_{c}$, out of energy eigenvalue function which is given in Eq.\eqref{energyc2}. We find
\begin{equation}\label{omegacrit2}
\begin{split}
		\tilde{\omega}_{c}=&\frac{1}{8\left(\ell-\beta k\right)^2}\left\lbrace\frac{1}{\mathrm{m}}\bigg(4\left(k^2+2\mathrm{m}C_{3}\right)\left(\ell-\beta k\right)+8\mathrm{M}\lambda\left(2\mathrm{m}C_{2}+\left(1+2n\right)^2\right) \right.\\
		 & \left.+16\mathrm{M}\lambda\right.\left(1+2n\right)\sqrt{2\mathrm{m}C_2+\left(\ell-\beta k\right)^2}\bigg)\left.\pm\left[\frac{1}{m^2}\Big(4\left(k^2+2\mathrm{m}C_{3}\right)\left(\ell-\beta k\right)\right.\right.\\
		&\left.\left.+8\mathrm{M}\lambda\left(2\mathrm{m}C_{2}+\left(1+2n\right)^2\right)+16\mathrm{M}\lambda\left(1+2n\right)\big(2\mathrm{m}C_2+\left(\ell-\beta k\right)^2\big)^{1/2}\Big)^2\right.\right.\\
		&\left.\left.+\frac{16}{m^2}\left(\ell-\beta k\right)^2\bigg(8mC_{1}\left(\big(\ell-\beta k\big)^2+2\mathrm{m}C_{2}+\left(1+2n\right)^2\right)-\left(k^2+2\mathrm{m}C_{3}\right)^2\right.\right.\\
		&\left.\left.+4\mathrm{M}\lambda\left(k^2+2\mathrm{m}C_{3}\right)\left(\ell-\beta k\right)+4\mathrm{M}^2\lambda^2\left(2\mathrm{m}C_2+\left(1+2n\right)^2\right)\right.\right.\\
		&\left.\left.+8\left(2\mathrm{m}C_{1}+\mathrm{M}^2\lambda^2\right)\left(1+2n\right)\left(2\mathrm{m}C_2+\left(\ell-\beta k\right)^2\right)^{1/2}\bigg)\right]^{1/2}\right\rbrace,
	\end{split}
\end{equation}
Alike the first case, we take the positive sign in the above equation. Then, we assume the following parameters for the numerical calculations: $\ell=\lambda=\mathrm{m}=\mathrm{M}=C_{1}=C_{2}=C_{3}=1$, and $k=0.5$. From Eq.\eqref{omegacrit2}, we calculate the critical angular velocity values for the ground and the first excited states as $26.41$, $77.71$, $157.39$ and $265.50$, respectively. If we consider the scalar potential energy  as consisted only of the radial harmonic term, that is $C=1$ and $C_2=C_3=0$, we find the following critical angular velocity values for the first four states: $10.69$, $49.82$, $117.39$ and $213.39$. Instead, if we consider solely the inverse square energy as the scalar potential energy, then we find $22.88$, $74.09$, $153.75$, and $261.85$. It is worth noting that, when $C_1=C_2=C_3$ are taken as zero, we obtain the same results of the first case.

\begin{figure}[!htb]
	\begin{center}
\includegraphics[scale=1]{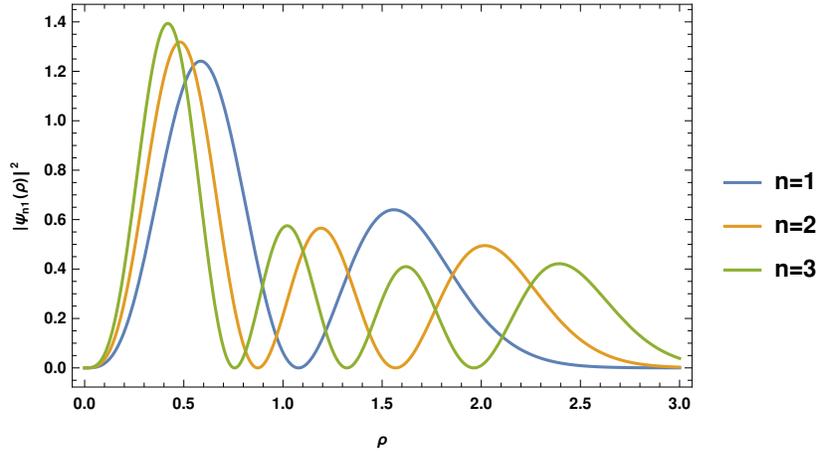}
		\caption{Three probability density functions versus  $\rho$. }\label{Fig1}
	\end{center}
\end{figure}

In order to examine the highest probability of finding a particle, we plot the probability density  functions, $\left|\psi_{n\ell}\left(\rho\right)\right|^2$, versus $\rho$ in Fig.~\ref{Fig1}.  Therefore, at first, we find there different normalized wave-functions from Eq.\eqref{WF3} for  $\beta=0.5$, where we have calculated the normalization coefficients of the first three wave functions as $2.10$, $1.62$, and $1.35$.  From Fig.~\ref{Fig1},  we observe that closest peaks to the origin  have the largest amplitudes.

\begin{figure}[!htb]
	\begin{center}
		\includegraphics[scale=1]{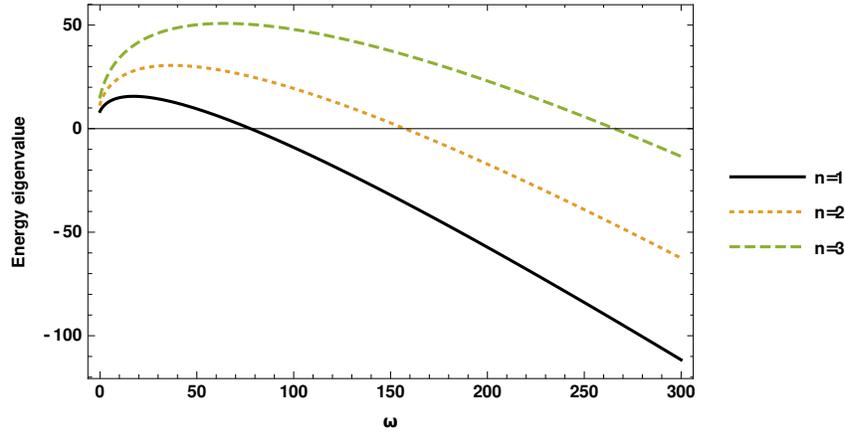}
		\caption{First three excited state's eigenvalue functions versus the angular velocity.}\label{Fig2}
	\end{center}
\end{figure}

To illustrate the effect of the constant angular velocity of the rotating frame on the energy eigenvalues, we plot  Fig. \ref{Fig2}. Here, we consider the first three excited states, thus, by using Eq. \eqref{energyc2} we plot  $\mathcal{E}_{11}$, $\mathcal{E}_{21}$, and $\mathcal{E}_{31}$ versus the angular velocity for $\beta=0.5$.  In Fig. \ref{Fig2}, initially, we observe a rise in the energy eigenvalues for the increasing  values of $\omega$, but after a certain  value of $\omega$, we see a decrease in the energy eigenvalues. We calculate the critical angular velocities as $77.71$, $157.39$, and $265.50$, respectively. We observe that these results are in the line with the plots. We see that for confinement, the relatively higher excited states need higher critical angular velocity values. This conclusion is also valid for the solely radial harmonic and inverse square cases, where the first three excited states' critical angular velocities are $49.82$, $117.39$, $213.39$ and $74.09$, $153.75$, $261.85$, respectively.

\begin{figure}[!htb]
\begin{center}
\includegraphics[scale=1]{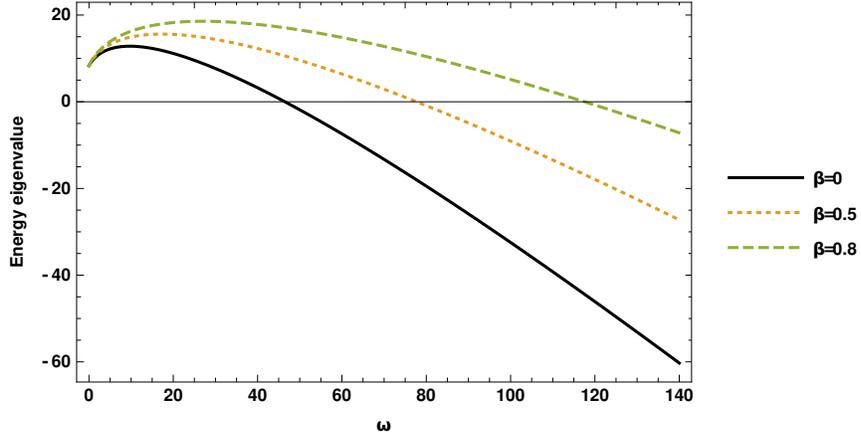}
\caption{The energy eigenvalue function for the first excited state of the pseudo-harmonic scalar potential energy case via three different Burger parameters. }\label{Figy3}
\end{center}
\end{figure}

\begin{figure}[!htb]
\begin{center}
\includegraphics[scale=1]{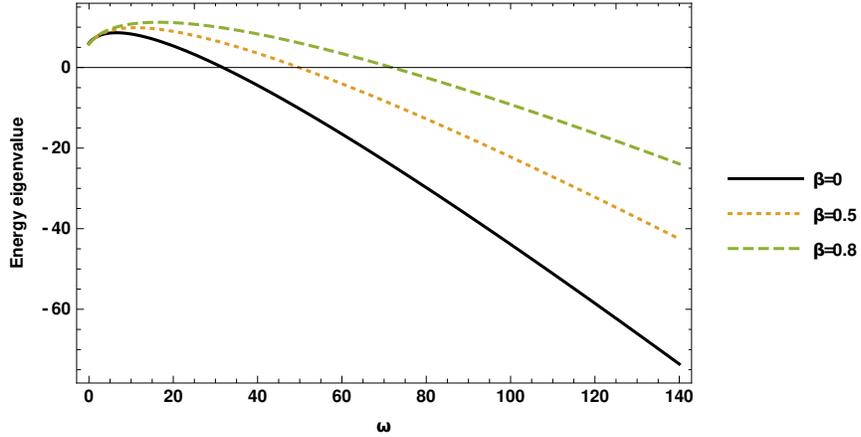}
\caption{The energy eigenvalue function for the first excited state of the harmonic scalar potential energy case via three different Burger parameters.}\label{Figy4}
\end{center}
\end{figure}

\begin{figure}[!htb]
\begin{center}
\includegraphics[scale=1]{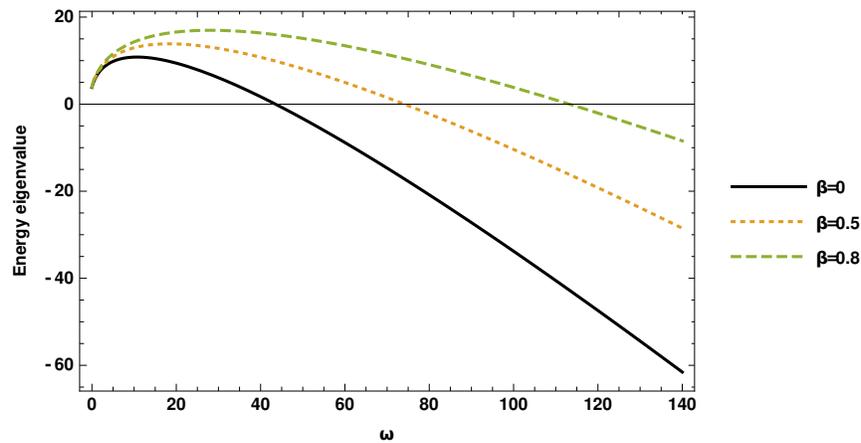}
\caption{The energy eigenvalue function for the first excited state of the inverse square scalar potential energy case via three different Burger parameters.}\label{Figy5}
\end{center}
\end{figure}

Next, we investigate the effect of the dislocation of the lattice on the critical angular velocity and  energy eigenvalues. We consider the first excited state of the pseudo-harmonic, harmonic and inverse square cases, and depict their energy eigenvalue functions versus the angular velocity with three Burger parameters in Figs.~\ref{Figy3}, \ref{Figy4} and \ref{Figy5}, respectively. In Fig.~\ref{Figy3}, in other words on the pseudo-harmonic scalar potential energy case, the increase of the dislocation parameter modifies the critical angular velocity values. In the absence of deformation, we calculate the angular velocity as $46.48$. This value arises to $117.52$  when  the dislocation parameter is equal to $0.8$. After this angular velocity, the particle becomes confined.

We observe the same characteristic behavior for the other potential energy forms. For example, in the harmonic energy and inverse square cases, we obtain the following critical angular velocity values  $31.74$, $49.82$, $71.91$; and $43.53$, $74.09$, $113.24$  for the Burger parameters of $0$, $0.5$ and $0.8$, respectively.

%Finally, we demonstrate the effect of the magnetic quadrupole parameter on the energy eigenvalue function. In Fig. \ref{Fig3}, we plot the first excited states' energy eigenvalue function versus $\mathrm{M}$ with three different dislocation parameters. Here, we have employed a random constant  angular velocity value, i.e. $\omega=100$. We observe that the energy eigenvalue increases as M increases. Up to a critical magnetic quadrupole parameter the particle is confined. We find this critical value as $2.45$ for the absence of the dislocation. For the non-zero Burger parameter appears this critical magnetic quadrupole value starts to decrease. For example for $\beta=0.5$ and $\beta=0.8$ it reduces to $1.31$ and $0.84$, respectively. We conclude that when the disorder increases, the particle's confinement becomes harder.

%\begin{figure}[!htb]
%\begin{center}
%\includegraphics[scale=1]{En21vM.eps}
%\caption{The energy eigenvalue function for the %first excited state versus the magnetic %quadrupole moment.}\label{Fig3}
%\end{center}
%\end{figure}

\section{Conclusion\label{Conc}}
In this study, we examine the interaction of a magnetic quadrupole moment of a moving particle in an elastic medium with a rotating frame in the presence of a screw dislocation. This medium possesses a radial electric field, therefore non-relativistic particle is influenced by a uniform effective magnetic field. We focus on the magnetic quadrupole polarizability tensor and solve the Schr\"odinger equation  exactly  by using the NU method for two particular cases. In the first case, we assume that the interaction in the absence of the static scalar potential energy. Then, in the second case, we consider a pseudo-harmonic oscillator like  static  potential energy. We acquire the wave-function and energy eigenvalues for both cases.  In both cases, we find that the dislocation parameter modifies the energy eigenvalue functions. We show that the particle confinement is based on the critical angular velocity value which is not a constant and depends on the dislocation parameter.
We present and discuss the behavior of energy eigenvalues versus the critical angular velocity with different Burger parameters in other forms of the scalar potential energies. Moreover, in the first case we examine the role of the dislocation and the angular velocity parameter on the degeneracy of the energy eigenvalues. We find that angular velocity removes the degeneracy of the system, while the quantum number $\ell$ effects the degeneracy. In the second case, we demonstrate and discuss the probability density and energy eigenvalues functions that correspond to the wave-function versus angular velocity parameter.
%and magnetic quadrupole moment parameter.

\section*{Acknowledgment}
The authors thank the referees for a thorough reading of our manuscript and for constructive suggestion. This work is financially supported from the Excellence project of Faculty of Science of University of Hradec Kr\'{a}lov\'{e}, $[2020/2209]$.  %One of the author, B.C. L\" utf\"uo\u{g}lu, was partially supported by the Turkish Science and Research Council (T\"{U}B\.{I}TAK).

%%%%%%%%%%%%%%%%%%%%%%%%%%%%%%%%%%%%%%%%%%%%%%%%

%% REFERENCES

%%%%%%%%%%%%%%%%%%%%%%%%%%%%%%%%%%%%%%%%%%%%%%%%

\end{document}